\documentclass[a4paper,12pt]{article}
\usepackage[english]{babel}
\usepackage{amsmath,amsfonts,amssymb,bm}
\usepackage{feynmp-auto}
\usepackage[numbers,comma,square,compress]{natbib}
\usepackage[margin=28mm]{geometry}
\usepackage[colorlinks=true,allcolors=blue]{hyperref}

\newcommand{\email}[1]{\href{mailto:#1}{#1}}

\begin{document}
\begin{center}
{\Large{\bf Can the gamma-ray bursts travelling through the interstellar
space be explained without invoking the drastic assumption
of Lorentz invariance violation?}}
\end{center}

\vspace*{.5em}
\begin{center}
{\bf
{M.~Chaichian$^{a,}$\footnote{\email{masud.chaichian@helsinki.fi}},
 I.~Brevik$^{b,}$\footnote{\email{iver.h.brevik@ntnu.no}} and
M.~Oksanen$^{a,}$\footnote{\email{markku.oksanen@helsinki.fi}}}}
\end{center}

\vspace*{.3em}
\begin{center}
{\it $^a$Department of Physics, University of Helsinki, P.O.Box 64,
\\FI-00014 Helsinki, Finland\\
$^b$Department of Energy and Process Engineering, Norwegian
University of Science\\ and Technology, N-7491 Trondheim, Norway}
\end{center}

\begin{abstract}
Experimental observations indicate that gamma-ray bursts (GRB)
and high-energy neutrino bursts may travel at different speeds with a
typical delay measured at the order of hours or days. We discuss two
potential interpretations for the GRB delay: dispersion of light in
interstellar medium and violation of Lorentz invariance due to quantum
gravitational fluctuations. Among a few other media, we consider
dispersion of light in an axion plasma, obtaining the axion plasma
frequency and the dispersion relation from quantum field theory for the
first time. We find that the density of axions inferred from
observations is far too low to produce the observed GRB delay. However,
a more precise estimation of the spatial distribution of axions is
required for a conclusive result. Other known media are also unable to
account for the GRB delay, although there remains uncertainties in the
observations of the delays. The interpretation in terms of Lorentz
invariance violation and modified dispersion relation suffers from its
own problems: since the modification of the dispersion relation should
not be dependent on particle type, delays between photons and neutrinos
are hard to explain. Thus neither interpretation is sufficient to
explain the observations. We conclude that a crucial difference between
the two interpretations is the frequency dependence of the propagation
speed of radiation: in dispersive plasma the group speed increases with
higher frequency, while Lorentz invariance violation implies lower speed
at higher frequency. Future experiments shall resolve which one of the
two frequency dependencies of GRB is actually the case.
\end{abstract}

\vspace*{2em}
\noindent
\emph{Talk given in 40th International Conference on High Energy Physics
- ICHEP2020\\
July 28 -- August 6, 2020\\
Prague, Czech Republic (virtual meeting)}

\pagebreak

\section{Introduction}
Gamma-ray bursts (GRB) are highly energetic and diverse events, which
are thought to be produced by violent stellar processes, in particular
supernovas and mergers of binary neutron stars. Those events may also
produce high-energy cosmic rays and consequently bursts of high-energy
neutrinos \cite{Waxman:1997ti}. Neutrino bursts have been observed to be
shifted in time with respect to the GRB (see
\cite{Adrian-Martinez:2016xij,Aartsen:2020eof} and references therein).
The time window $\tau=t_{GRB}-t_{\nu}$ between the arrival times of a
GRB $t_{GRB}$ and a neutrino burst $t_{\nu}$ can vary between an hour or
several days. Assuming that a GRB and the corresponding neutrino burst
are produced at the same time or within a short period, a significant
delay $\tau$ would indicate that the electromagnetic and neutrino
signals have travelled at different speeds. Note, however, that the
recent experimental studies show only faint neutrino signals associated
with GRB \cite{Adrian-Martinez:2016xij,Aartsen:2020eof}, and hence the
observed delays may be inaccurate.

It is equally challenging to interpret the GRB delay within standard
physics where GRB are delayed due to the interaction of photons with
interstellar media, a phenomenon which always occurs.
In this way, one can also shed additional light on the
``microstructure'' of the Universe or a part of it and its
constituents. The interaction of neutrinos with any interstellar medium
is extremely weak and hence the dispersion of neutrinos is negligible.
Secondly, while the neutrinos are massive and oscillating, the effect on
the speed of high-energy neutrinos is very small. Consider a GRB with
photon energy 1 TeV and neutrinos with the same energy,
$E=1\;\mathrm{TeV}$. The dispersion relation $E^2=p^2c^2+m^2 c^4$ gives
the speed of the neutrinos as $v_\nu=\frac{dE}{dp}\approx c(1-d_\nu)$,
where $d_\nu=\frac{m^2c^4}{2E^2}$. Averaging over 3 neutrinos, $\langle
m^2c^4\rangle = (1/3) (0.1\,\mathrm{eV})^2$, where the masses are
estimated with the heaviest neutrino mass. The speed of neutrinos is
given by $d_\nu=0.17 \times 10^{-26}$. Thus the delay compared to a
signal travelling at the speed $c$ would be measured in nanoseconds even
for signals from furthest galaxies: $\tau=D\times
d_\nu/c\lesssim10^{-8}\;\mathrm{s}$, using a maximal travelling distance
$D=10^{27}\;\mathrm{m}$ (across the whole universe). That is negligible
compared to the observed GRB delays. Theories of neutrino production in
GRB actually predict neutrinos with even higher energy of order
$10^{2}$--$10^{7}$ TeV \cite{Waxman:1997ti}, which means $v_{\nu}$ is
even closer to $c$. This justifies $v_{\nu}=c$ in our estimates.

Violation of Lorentz invariance and the associated modification of the
dispersion relation has been considered as a potential interpretation
of the delay of high-energy GRB
\cite{AmelinoCamelia:1997gz,Jacob:2006gn}.
This approach is motivated by various approaches to quantum gravity,
since quantum-gravitational fluctuations may lead to a non-trivial
refractive index \cite{Ellis:2008gg}. We shall comment the Lorentz
invariance violation interpretation in Sec.~\ref{LIV}.

Before seeking to modify fundamental principles such as Lorentz
invariance we prefer to consider possible explanations for the observed
phenomena by means of standard physics.
We consider the dispersion of light in several media and assess the
produced GRB delay when photons and neutrinos are assumed to
be emitted from the same source at the same time. Neither electron
plasma nor photon plasma can account for the observed GRB delay.
Then we consider axions. Axions are pseudoscalar particles that may both
provide a solution to the strong CP problem and constitute cold dark
matter. Axions are not electrically charged, since a charged axion would
be luminuous, but can still interact with photons. Axion electrodynamics
has been studied actively and it is connected to topological insulators
\cite{Wilczek:1987mv,Martin-Ruiz:2015skg}. Therefore an axion plasma is
a plausible cosmic medium that would have an effect on the propagation
of light from distant galaxies. We derive the dispersion relation in an
axion plasma and assess its effect on the GRB delay.

\section{Dispersion relation and plasma frequency}
A plasma can support both longitudinal and transverse waves.
We are interested in transverse waves.
Dispersion relation for light in a plasma is
\begin{equation}\label{disprel}
\omega^2=c^2\bm{k}^2+\omega_p^2,
\end{equation}
where $\omega_p$ is the plasma frequency. The angular frequency is
also given as
$\omega=\bm{v}(\hat{\bm{k}})\cdot\bm{k}=\frac{c|\bm{k}|}{n}$,
where $n$ is the refraction index and $\bm{v}=\frac{c}{n}\hat{\bm{k}}$
is the phase velocity, where
$\hat{\bm{k}}=\frac{\bm{k}}{|\bm{k}|}$. Thus the refraction index is
related to the plasma frequency as
\begin{equation}\label{n^2}
n^2=1-\frac{\omega_p^2}{\omega^2}.
\end{equation}
In an isotropic medium, $\omega=\omega(|\bm{k}|)$ and
$\omega_p=\omega_p(|\bm{k}|)$, group velocity is parallel to phase
velocity.
When the photon momentum is large compared to the plasma frequency,
$c^2\bm{k}^2\gg\omega_p^2$, we obtain that group velocity is
only slightly lower than $c$,
\begin{equation}\label{v_g.plasma}
v_g=\frac{\partial\omega(|\bm{k}|)}{\partial|\bm{k}|}
\simeq c\left( 1-\frac{\omega_p^2}{2\omega^2} \right)\equiv c(1-d).
\end{equation}

Now we explain how the plasma frequency and refraction index can be
derived from quantum field theory.
From here on we assume units $\hbar=c=\epsilon_0=1$.
The refraction index is related to the forward scattering amplitude
$f(0)$ as \cite{Bohr-Peierls-Placzek}
\begin{equation}\label{n.qm}
n=1+2\pi\frac{Nf(0)}{\omega^2},
\end{equation}
where $N$ is the number density of scatterers. The relation \eqref{n.qm}
is valid when $n$ is close to one, $|n-1|\ll1$, and follows from the
inteference between incident and scattered waves.
Inserting  \eqref{n^2} into \eqref{n.qm}, we obtain a relation between
the plasma frequency and the scattering amplitude.
When the photon frequency is large compared to the plasma frequency,
$\omega^2\gg\omega_p^2$, the relation is given as
\begin{equation}\label{omega_p^2}
\omega_p^2=-4\pi Nf(0).
\end{equation}
The scattering amplitude $f(\theta)$ is defined as a part of the
wavefunction at large distance $r$ from the scatterer,
$\psi(\bm{r})=C\left(e^{i\bm{k}\cdot\bm{r}}+f(\theta)e^{ikr}/r\right)$,
where $C$ is a normalization factor. The differential cross section is
given in terms of the scattering amplitude as
$d\sigma(\theta)=|f(\theta)|^2d\Omega$.
The differential cross section can as well be obtained from quantum
field theory. Hence we obtain the differential cross section $d\sigma$
at angle $\theta=0$ in quantum field theory and identify the forward
scattering amplitude as
\begin{equation}\label{fsa.abs}
|f(0)|=\left( \frac{d\sigma(0)}{d\Omega} \right)^{\frac{1}{2}}.
\end{equation}

For an electron plasma, we obtain the differential cross section for
scattering of a photon on an electron in quantum field theory. In the
rest frame of the initial electron, we obtain
$d\sigma(0)=(e^4/16\pi^2m_e^2)d\Omega$, and according to
\eqref{fsa.abs} we get $|f(0)|=\alpha/m_e$,where $\alpha$ is the
fine-structure constant, $\alpha=e^2/4\pi$, and $m_e$ is the electron
mass. Thus the plasma frequency $\eqref{omega_p^2}$ is given as
\begin{equation}\label{pf.electron}
\omega_p^2=\frac{Ne^2}{m_e},
\end{equation}
which is the same result that is obtained from classical electrodynamics
\cite{CMRT-ED,Jackson}.
With $m_e = 0.511~$MeV, we obtain the group velocity \eqref{v_g.plasma}
for the photon energy 1 TeV,
\begin{equation}
v_{GRB}=c(1-0.7\times 10^{-51}\times N\times\mathrm{metre}^3).
\end{equation}
Therefore, dispersive properties of an electron gas are not significant
enough to account for a time delay of the order of several hours as
observed.

Dispersion of light in light plasma also produces a too small delay. We
obtain from light on light scattering $\bm{\omega}_p^2=
\mathrm{const.}\times N_{\gamma}e^4/\omega$, where
$N_{\gamma}$ is the number density of photons. For a delay of the
order of few hours, we would need photon density
$N_{\gamma}=10^{39}\;\mathrm{m}^{-3}$, while according to the Planck
data on the CMB (Cosmic Microwave Background) radiation:
$N_{\gamma}=(4$--$5)\times10^{8}\;\mathrm{m}^{-3}$.

\section{Axion plasma and its effect on the propagation of gamma-rays}

Interaction Lagrangian of axion electrodynamics is
\cite{Sikivie:2020zpn}
\begin{equation}
\mathcal{L}_{a\gamma\gamma}=-\frac{1}{4}ga F_{\mu\nu}\tilde{F}^{\mu\nu}
=-\frac{1}{2}ga\epsilon^{\mu\nu\rho\sigma}\partial_\mu A_\nu
\partial_\rho A_\sigma,
\end{equation}
where $g$ is a coupling constant, $a$ is the axion pseudoscalar field,
the electromagnetic field strength tensor is
$F_{\mu\nu}=\partial_\mu A_\nu-\partial_\nu A_\mu$, and its dual
$\tilde{F}^{\mu\nu}=\frac{1}{2}\epsilon^{\mu\nu\rho\sigma}
F_{\rho\sigma}$.

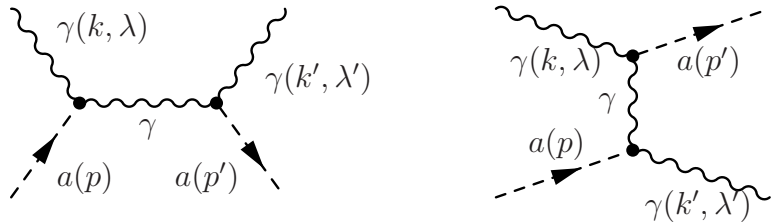
\begin{figure}[ht]
\begin{center}
\begin{fmffile}{axion-diagrams}
\begin{fmfgraph*}(45,25)
\fmfleft{i1,i2} \fmfright{o1,o2}
\fmf{scalar,label=$a(p)$}{i1,v1}
\fmf{photon,label=$\gamma(k,,\lambda)$}{i2,v1}
\fmf{photon,label=$\gamma$}{v1,v2}
\fmf{scalar,label=$a(p')$}{v2,o1}
\fmf{photon,label=$\gamma(k',,\lambda')$}{v2,o2}
\fmfdot{v1,v2}
\end{fmfgraph*}
\hspace{16mm}
\begin{fmfgraph*}(45,25)
\fmfleft{i1,i2} \fmfright{o1,o2}
\fmf{scalar,label=$a(p)$}{i1,v2}
\fmf{photon,label=$\gamma(k,,\lambda)$}{i2,v1}
\fmf{photon,label=$\gamma$}{v1,v2}
\fmf{photon,label=$\gamma(k',,\lambda')$}{v2,o1}
\fmf{scalar,label=$a(p')$,label.side=right}{v1,o2}
\fmfdot{v1,v2}
\end{fmfgraph*}
\end{fmffile}
\end{center}
\caption{Feynman diagrams for scattering of photon on axion
(drawn from left to right)}
\label{fig1}
\end{figure}

Consider scattering of photon on axion $\gamma+a\rightarrow\gamma+a$ at
tree level. The scattering amplitude $\mathcal{M}$ is a sum of two terms
represented by the diagrams in Fig.~\ref{fig1}.
\unitlength = 1mm 
We are interested in scattering with parallel momenta for initial and
final photons, i.e. with angle $\theta=0$. The differential cross
section for unpolarized photons with angle $\theta=0$ is obtained in
the rest frame of the initial axion as
\begin{equation}\label{dcs.axion}
d\sigma(0)=\frac{1}{64\pi^2m_a^2}
\left( \frac{1}{2}\sum_{\lambda,\lambda'}|\mathcal{M}(0)|^2 \right)
d\Omega
=\left( \frac{3g^2}{16\pi} \right)^2 \left|
\frac{\omega^2}{(2\omega+m_a)}
+\frac{\omega^2}{(-2\omega+m_a)} \right|^2 d\Omega.
\end{equation}
where $\omega$ is the energy of the initial photon.
The scattering amplitude $|f(0)|$ is obtained according to
\eqref{fsa.abs}.
Since the axions are very light, $m_a\sim10^{-5}\;$eV, and we are
interested in very high energy photons, we consider the limit
$\omega\gg m_a$. For photons with energies well above the axion mass,
the plasma frequency is nearly constant, i.e. independent of the
frequency of the incoming light:
\begin{equation}\label{hepf.axion}
\omega_p^2=\frac{3}{8}Ng^2m_a\left( 1+\frac{m_a^2}{4\omega^2}
+\mathcal{O}\left(\frac{m_a^4}{\omega^4}\right) \right)
\simeq\frac{3}{8}Ng^2m_a.
\end{equation}

Estimating the effective coupling constant to be
$g=10^{-10}\;\mathrm{GeV}^{-1}$, and the axion mass $m_a= 10^{-5}\;$eV,
we obtain the group velocity for the photon of energy 1 TeV in an
axion plasma,
\begin{equation}
v_{GRB}= c(1-d), \quad
d=\frac{3}{16}\frac{Ng^2m_a}{\omega^2}
=1.4\times 10^{-88}\times N\times \mathrm{metre}^{3}.
\end{equation}
Thus the delay of the GRB in the galactic plasma is
\begin{equation}
\tau= \frac{D\times d}{c}=
\frac{3}{16}\frac{Ng^2m_a}{\omega^2}\frac{D}{c}
=4.8\times 10^{-97}\times D\times N\times\mathrm{metre}^{2}
\times\mathrm{second},
\end{equation}
where $D$ is the distance traveled by the photons. Typical value of the
delay $\tau$ taken from ANTARES data is $\tau=3.25$ hours.
The effective distance travelled by photons in expanding Universe
depends on the redshift $z$ \cite{Adrian-Martinez:2016xij},
$D(z)=\frac{c}{H_0}\int_0^z\frac{(1+z)dz}{\sqrt{\Omega_m(1+z)^3
+\Omega_\Lambda}}$. If $D$ is taken as the diameter of observable
Universe, $D = 8.8 \times 10^ {26}$ m, we need an axion number density
$N\simeq10^{73}\;\mathrm{m}^{-3}$. This is a very large number density
that apparently contradicts experimental data.

A more realistic scenario is to consider that axions are concentrated
in galactic halos (constituting cold dark matter). The mass density of
axions in a galactic halo is estimated $D_m=0.45\;\mathrm{GeV/cm}^3$,
and the radius of the halo is $5\times 10^{20}\;\mathrm{m}$
\cite{Braine:2019fqb}. Number density of axions is $N_{GH}=0.45\times
10^{20}\; \mathrm{m}^{-3}$. In order to produce the delay $\tau=3.25\;
\mathrm{hours}$, the axion number density in galactic halo should be
$N\simeq10^{79}\;\mathrm{m}^{-3}$, which is much higher than from data
$N_{GH}$ multiplied by any number of farther galaxies within the
diameter of the Universe.

Details of the derivation of \eqref{hepf.axion} will be presented
elsewhere \cite{calculation}.
For related works on the bending of light in axion backgrounds but not
considering the issues concerning GRB, see
\cite{McDonald:2020why} and references therein.

\section{Resolution between dispersion in plasma and Lorentz invariance
violation}
\label{LIV}

In the quantum gravity motivated interpretation that violates Lorentz
invariance \cite{AmelinoCamelia:1997gz,Jacob:2006gn}, the dispersion
relation is modified to contain higher-power energy terms (or
higher-power momentum terms),
$E^2[1+\sum_{n=1}\xi_n(E/E_{QG})^n]=p^2c^2+m^2c^4$.
Then the group velocity of light is
\begin{equation}\label{v_g.LIV}
v_g=c\left( 1-\xi\frac{E}{E_{QG}}
+\mathcal{O}\left(\frac{E^2}{E_{QG}^2}\right) \right),
\end{equation}
where $E_{QG}$ is an effective quantum gravity energy scale, usually of
order $E_{QG}=10^{16}\;$GeV.
Hence the modification of the dispersion relation implies that the
slowdown of radiation is increased with higher energy. Thus this
approach has mainly been used to consider the delay between higher
energy photons and lower energy photons produced in GRB. A
delay between neutrinos and photons produced in GRB might
be possible in this interpretation only if the energy of neutrinos is
several orders of magnitude higher than the energy of photons
\cite{Jacob:2006gn}.

The key feature that differentiates the dispersive plasma
interpretation from the Lorentz invariance violation (LIV)
interpretation is the energy dependence of the signal delay. In plasma
the delay decreases with higher photon frequency,
$\tau\propto\omega^{-2}$, while in the LIV case it increases with
frequency, $\tau\propto\omega$. It would be crucial to test the
frequency/energy dependence of the delay experimentally.
That requires the exact measurement time of observation of GRB and
spectral resolutions and therefore, the planned broad energy range
measurements are utmost crucial \cite{Acciari:2019dbx}.

Since neither LIV nor dispersion of light in a plasma can explain such a
large delay between GRB and neutrinos, one could even suspect the
existence of the delay with such an amount.

\newcommand{\atitle}[1]{\emph{#1},}
\newcommand{\jref}[2]{\href{https://doi.org/#2}{#1}}
\newcommand{\arXiv}[2]{\href{http://arxiv.org/abs/#1}
{\texttt{arXiv:#1 [#2]}}}
\newcommand{\arXivOld}[1]{\href{http://arxiv.org/abs/#1}{\texttt{#1}}}

\end{document}